\documentclass[twocolumn,bibnotes]{revtex4}

\usepackage{color,rotating,graphicx}
\usepackage{amsmath,ulem,comment}
\usepackage{dsfont}

\newcommand{\ri}{{\rm i}}
\newcommand{\re}{{\rm e}}
\newcommand{\rd}{{\rm d}}

\begin{document}

\title{Enhancement of synthetic magnetic field induced nonreciprocity via bound states in continuum in dissipatively coupled systems}

\author{S.-A. Biehs}
\email{s.age.biehs@uni-oldenburg.de}
\affiliation{Institut f\"{u}r Physik, Carl von Ossietzky Universit\"{a}t, D-26111 Oldenburg, Germany}
\affiliation{Center for Nanoscale Dynamics  (CeNaD), Carl von Ossietzky Universit\"{a}t, D-26129 Oldenburg, Germany}
\affiliation{Laboratoire Charles Coulomb (L2C), UMR 5221 CNRS-Université de Montpellier, F-34095 Montpellier, France}
\author{G. S. Agarwal}
\email{girish.agarwal@tamu.edu}
\affiliation{ Institute for Quantum Science and Engineering and Department of Biological and 
Agricultural Engineering Department of Physics and Astronomy, Texas A \& M University, College Station, Texas 77845, USA}

\begin{abstract}
The nonreciprocal propagation of light typically requires use of materials like ferrites or magneto-optical media with a strong magnetic bias or methods based on material nonlinearities which require use of strong electromagnetic fields. A simpler possibility to produce nonreciprocity is to use spatio-temporal modulations to produce magnetic fields in synthetic dimensions. In this paper we show that dissipatively coupled systems can lead to considerable enhancement of nonreciprocity in synthetic fields. The enhancement comes about from the existence of nearly nondecaying mode -bound state in continuum (BIC) in dissipatively coupled systems. The dissipative coupling occurs in a wide class of systems coupled via transmission lines, waveguides, or nano fibers. The systems could be optical resonators or microscopic qubits. Remarkably we find that for specific choice of the modulation amplitudes, the transmission say in forward direction is completely extinguished whereas in the backward direction it becomes maximum. The synthetic fields produce transmission resonances which show significant line narrowing which owe their origin to existence of BIC’s in dissipative systems.
\end{abstract}

\maketitle

\section{Introduction}

%general discussion of nonreciprocity
The control of photonic energy and heat transport is typically relying on nonreciprocal light-matter interactions which can be achieved in different ways. One possibility is to employ nonreciprocal media such as ferro-magnets, ferrits, and magneto-optical media which need a magnetic bias to obtain a time-reversal symmetry breaking~\cite{Alu}. The nonreciprocity manifests itself in many different ways. To have just two out of many examples, there can be a highly non-reciprocal energy transport~\cite{Doyeux2017} or heat transport~\cite{NRdiode} by coupling of two-level systems or two nanoparticles to the nonreciprocal surface waves of a nearby medium or substrate. The latter effect can also be understood by the spin-spin coupling~\cite{SpinSpin} of particle resonances with the spin-momentum locked surface waves of the nearby sample~\cite{Zubin2019} and hence, the mechanism is reminiscent of the chiral emission of quantum emitters~\cite{LodahlEtAl2017}. Similar, nonreciprocal effects can also be obtained with intrinsically nonreciprocal materials such as Weyl semi-metals. The advantage is that it is not necessary to apply any external magnetic field to obtain for example nonreciprocal reflection~\cite{ZhaouEtAl2020,Pajovic2020} and therefore a violation of Kirchhoff's law, directional energy or heat transport~\cite{HuEtAl2023}, directional spin-controlled thermal emission~\cite{DongEtAl2021}, and anomalous Hall effect for thermal radiation~\cite{hall,ahall}. Even though Weyl semi-metals offer a route for nonreciprocal effects without external magnetic fields, they do not allow for an active control of the nonreciprocity as can be done by changing the magnitude or direction of the applied magnetic fields in magneto-optical systems, for instance. Since an integration of magnetic fields in nanoscale devices is not easy and the nonreciprocal effects are typically small, other methods for achieving nonreciprocity have been explored. 

%general modulation methods
One possibility is to utilize the phase matching conditions in nonlinear effects like Brillouin-scattering-induced transparency which can only be fulfilled in one direction~\cite{KimEtAl2015,DongEtAl2015}. Other approaches to achieve magnetic free nonreciprocity is to apply spatio-temporal modulations. For example, spatio-temporal modulations of permittivities or effective permittivities have been shown to result in a high degree of nonreciprocal energy transport in a wave-guide~\cite{LiraEtAl2012,ChamanaraEtAl2017} or coupled resonator-loop structure with angular-momentum biasing~\cite{EstepEtAl2014}. Such spatio-temporal modulations of permittivities can further result in violation of Kirchhoff's law due to the induced nonreciprocity of absorptivity and emissivity~\cite{GhanekarEtAl2022}. Similarly, specific modulation protocols around exceptional points permit nonreciprocal energy transfer as shown for Casimir force induced energy transfer~\cite{XuEtAl2020} and proposed for heat transfer in a three oscillator system~\cite{LiEtAl2019}.

%synthetic magnetic fields
A related method to achieve nonreciprocal energy transfer is to induce resonance frequency modulations which can be understood to be equivalent to applying electric and magnetic synthetic fields~\cite{Lustig}. Here, the synthetic electric field is induced by the frequency modulation which generates side-bands of the system resonances and the synthetic magnetic field for photons is generated by including phase-shifts in the modulation~\cite{FangEtAl2012,TzuanEtAl2014,OzawaEtAl2016}. Therefore, in the abstract synthetic space the magnetic field can break the time-reversal symmetry which results in nonreciprocal energy transport as theoretically and experimentally demonstrated~\cite{PetersonEtAl2019}. Corresponding to the violation of Kirchhoff's law~\cite{GhanekarEtAl2022} by a spatio-temporal modulation of the permittivity the presence of a synthetic magnetic field leads to a broken detailed balance for radiative heat exchange~\cite{SABGSA2023}. Interestingly, the heat flux is in this case reciprocal, but it can be nonreciprocal when specific combinations of frequency and coupling strength modulations in a three-resonator instead of two-resonator systems are carried out~\cite{AlcazarEtAl2021}.  

%dissipative coupling
%BIC

The aim of this work is to demonstrate enhancement of the nonreciprocal energy transfer in a system of two dissipatively coupled resonators with an external excitation by a plane electromagnetic wave and in presence of synthetic magnetic fields. In recent years the interest in dissipatively coupled systems has been growing due to their unique property that these can posses a mode which can have a very long lifetime~\cite{AzzamKildishev2021,YangEtAl2020,JayEtAl2021,GSAarxiv} analogous to a bound state in continuum (BIC). The unique enhanced sensing capabilities of such systems have been discussed in literature~\cite{JayEtAl2021,ZhangEtlAl2020}. Such systems have been used to demonstrate the effect of the phase dependent couplings on a number of physical effects such as electromagnetically induced transparency, nonreciprocal transport and interconversion between optical and microwave fields~\cite{Brongersma,NairEtAl2022,PengEtAl2016,ChoiEtAl2018,HarderEtAL2018}.  In this work we introduce the idea of synthetic magnetic fields in dissipatively coupled systems and discuss the nonreciprocal energy transport in such dissipatively coupled systems from a very general perspective. We bring out important differences between a dissipatively and a dispersively coupled system with a focus on the impact of the BIC and the nonreciprocal energy transport in presence of a synthetic magnetic field. The results of this study have applicability to a wide class of systems --- resonator physics, qubits coupled to fibers and waveguides, magnonics.

The article is organized as follows: In Sec.~II we introduce the dynamical equations of the two dissipatively and dispersively coupled oscillators pointing out the appearance of the BIC. In Sec.~III we introduce the modulation scheme as well as the second-order perturbation result for the transmission coefficients. In Sec.~IV we provide the formalism for the numerical exact evaluation of the transmission coefficient and discuss our numerical results. In Sec.~V we analyse the observed linewidth narrowing and we conclude with a summary in Sec.~VI. 

\section{Dissipative coupling and bound states in continuum}

\begin{figure}
  \includegraphics[angle=0,scale=0.45]{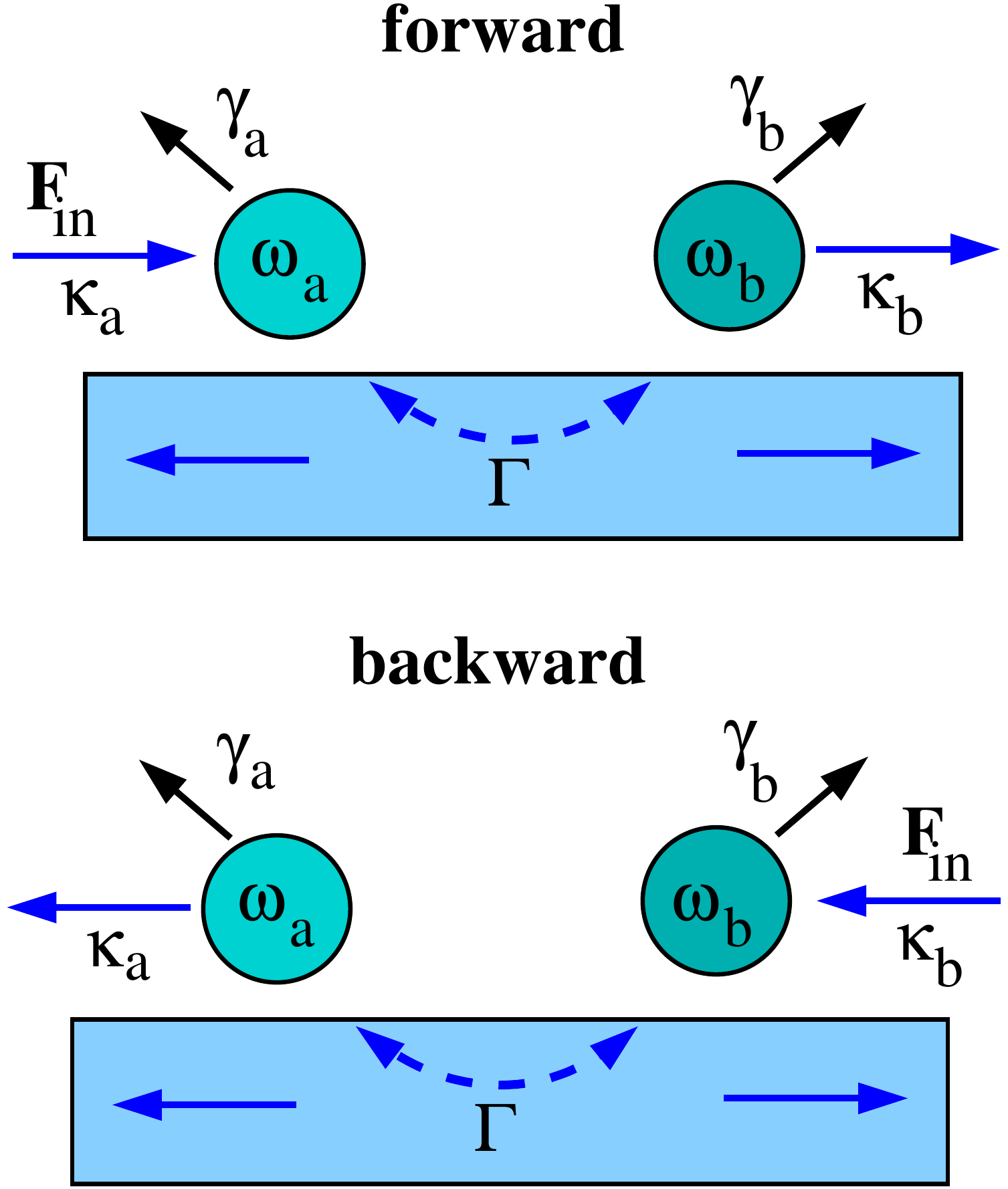}
	\caption{\label{Fig:DCcoupling} Sketch of the forward and backward transmission of an incident monochromatic field $\mathbf{F}_{\rm in}$ in a dissipative coupling scenario.}
\end{figure}

We consider now a dissipative coupling scenario as depicted in Fig.~\ref{Fig:DCcoupling} where two resonators are placed in close vicinity of a 1D waveguide with a monochromatic driving field exciting the oscillators. The master equation for $N$ resonators coupled by a shared 1D waveguide with linear dispersion in the Born-Markov approximation can be written as~\cite{MukhopadhyayEtAl2022}
\begin{equation}
\begin{split}
	\frac{\rd \rho_S(t)}{\rd t} &= - \frac{\ri}{\hbar} [H_S, \rho_S] - \ri \sum_{\alpha \neq \beta} \Omega_{\alpha\beta} [a_\alpha^\dagger a_\beta, \rho_S] \\
				    &\quad - \sum_{\alpha,\beta = 1}^N \kappa_{\alpha\beta} (a_\alpha^\dagger a_\beta \rho_S -2 a_\beta \rho_S a_\alpha^\dagger + \rho_S a^\dagger_\alpha a_\beta)
\end{split}
\end{equation}
where $\rho_S$ is the reduced density matrix of the system of resonators with resonance frequencies $\omega_\alpha$ described by the Hamiltonian $H_S = \hbar \sum_{\alpha = 1}^N \omega_\alpha a_\alpha^\dagger a_\alpha$. In general, the oscillators are coupled dispersively and dissipatively via the coupling constants given by~\cite{MukhopadhyayEtAl2022}.
\begin{align}
	\kappa_{\alpha\beta} &= \frac{g_\alpha^2 L}{v_g}\delta_{\alpha\beta} + \Gamma_{\alpha\beta} \cos(\phi_{\alpha\beta}) (1 - \delta_{\alpha\beta}), \\
	\Omega_{\alpha\beta} &= \sin(\phi_{\alpha\beta}) \Gamma_{\alpha\beta}.
\end{align}
Here, $v_g$ is the group velocity of the waveguide mode assuming a linear dispersion $\omega_k \approx v_g |k|$, $\phi_{\alpha\beta} = k_0 x_{\alpha\beta}$ is the phase acquired by the central waveguide mode ($k \approx k_0$) between two coupled resonators $\alpha$ and $\beta$ being separated by a distance $x_{\alpha\beta}$, and $\Gamma_{\alpha\beta} = g_\alpha g_\beta L/v_g$ with the length of the waveguide $L$. Obviously, the dispersive coupling constants $\Omega_{\alpha\beta}$ between two resonators $\alpha$ and $\beta$ vanish when choosing the relative distance $x_{\alpha\beta}$ between the resonators such that the product $k_0 x_{\alpha\beta}$ is an integral multiple of $\pi$. In this case the coupling is purely dissipative~\cite{MukhopadhyayEtAl2022}. If we now consider the case of two resonators with identical coupling to the waveguide $g_1 = g_2$, then $\Gamma_{11} = \Gamma_{12} = \Gamma_{21} = \Gamma_{22} \equiv \Gamma$. Furthermore, we add to the first loss term in the dissipative coupling constant internal and radiative losses $\gamma_{a/b}$ as well as couplings $\kappa_{a/b}$ to input and output ports so that $\kappa_{11} = \kappa_{22} = \Gamma + \gamma_{a/b} + \kappa_{a/b}$ and $\kappa_{12} = \kappa_{21} = \Gamma$. Then, the dynamical equation for the mean values $a = \langle a_1 \rangle$ and $b = \langle a_2\rangle$ is easily derived from the master equation we obtain    
\begin{equation}
	\frac{\rd}{\rd t} \begin{pmatrix} a \\ b\end{pmatrix} = \mathds{A}^{\rm dc}  \begin{pmatrix} a \\ b\end{pmatrix} + \mathbf{F}
\label{Eq:coupledOscillators}
\end{equation}
with the dynamical matrix
\begin{equation}
	\mathds{A}^{\rm dc} =  \begin{pmatrix}
                        	  -\ri \omega_a - (\gamma_a + \kappa_a + \Gamma) & -\Gamma \\
                        	  -\Gamma & -\ri \omega_b - (\gamma_b + \kappa_b + \Gamma) 
                               \end{pmatrix}.
\end{equation}
Here,  $\omega_{a/b}$ are the resonance frequencies of the two oscillators and $\mathbf{F}$ is an additional external driving field. The radiative and internal losses $\gamma_{a/b}$ are accompanied by the dissipation rate $\Gamma \in \mathds{R}$ which also serves as dissipative coupling constant between both resonators; $\kappa_{a/b}$ are the coupling constants to input or output ports. With our approach we ignore any quantum fluctuations which is reasonable since the resonators are driven by a classical field $\mathbf{F}$ which means that the photon numbers are large and quantum fluctuations are small. In the rotating frame of the driving field the dynamical matrix changes to
\begin{equation}
	\tilde{\mathds{A}}^{\rm dc}=  \begin{pmatrix}
		-\ri \delta_{a} - (\gamma_a + \kappa_a + \Gamma) & -\Gamma \\
				  -\Gamma & -\ri \delta_{b} - (\gamma_b + \kappa_b + \Gamma) 
                                  \end{pmatrix}.
\end{equation}
with detunings $\delta_{a/b} = \omega_{a/b} - \omega_l$ between the resonances and the driving field frequency $\omega_l$. As can be easily verified, the two eigenvalues in the limit $\gamma_a + \kappa_a, \gamma_b + \kappa_b \rightarrow 0$ with a zero detuning $\delta_{a}, \delta_{b} = 0$ are given by $\lambda_{A,+}^{\rm dc} = 2 \Gamma$ and $\lambda_{A,-}^{\rm dc} = 0$. Since the eigenvalue $\lambda_-$ is zero its eigenstate has the peculiar property that it does not evolve in time, i.e.\ it has an infinitely long lifetime. This is a clear signature of a BIC~\cite{AzzamKildishev2021} which does not exist in systems with pure dispersive coupling. 

In Fourier space the coupled mode equation (\ref{Eq:coupledOscillators}) can be written as
\begin{equation}
	\boldsymbol{\psi}^{(0)} = \mathds{M}^{\rm dc} \tilde{\mathbf{F}}
	\label{Eq:CoupledFourier}
\end{equation}
with the Fourier representations $\boldsymbol{\psi}^{(0)} = \bigl( a(\omega) , b(\omega) \bigr)^t$ and $\tilde{\mathbf{F}}$, and the matrix 
\begin{equation}
	\mathds{M}^{\rm dc} =  \frac{1}{X_a^{\rm dc} X_b^{\rm dc} - \Gamma^2}\begin{pmatrix} X_b^{\rm dc} & - \Gamma \\ - \Gamma & X_a^{\rm dc} \end{pmatrix}
\end{equation}
with
\begin{equation}
	X_{a/b}^{\rm dc} = \ri \delta_{a/b} + \gamma_{a/b} + \kappa_{a/b} + \Gamma
	\label{Eq9}
\end{equation}
and $\delta_{a/b} = \omega_{a/b} - \omega$. When focusing at the case where both oscillators have identical properties, i.e.\ $\omega_a = \omega_b \equiv \omega_0$ and $\gamma_a + \kappa_a = \gamma_b + \kappa_b \equiv \gamma$ and therefore $X_a^{\rm dc} = X_b^{\rm dc} \equiv X^{\rm dc}$, we find the eigenvalues $\lambda_{\pm}^{\rm dc} = (X^{\rm dc} \pm \Gamma)^{-1}$. Hence, when taking again the limit $\gamma \rightarrow 0$ with a zero detuning $\delta_a, \delta_b = 0$ we find
\begin{equation}
	\lambda_+^{\rm dc} = \frac{1}{\gamma + 2 \Gamma} \rightarrow  \frac{1}{2 \Gamma}
\end{equation}
and
\begin{equation}
	\lambda_-^{\rm dc} = \frac{1}{\gamma} \rightarrow \infty.
\end{equation}
Hence, in the frequency domain the BIC is characterized by an infinitely large eigenvalue of the matrix $\mathds{M}^{\rm dc}$ which means that for even for finite but small damping rates $\gamma$ the response to an external excitation by a driving field can be very strong with a linewidth proportional to $\gamma$. 

\begin{figure}
  \includegraphics[angle=0,scale=0.45]{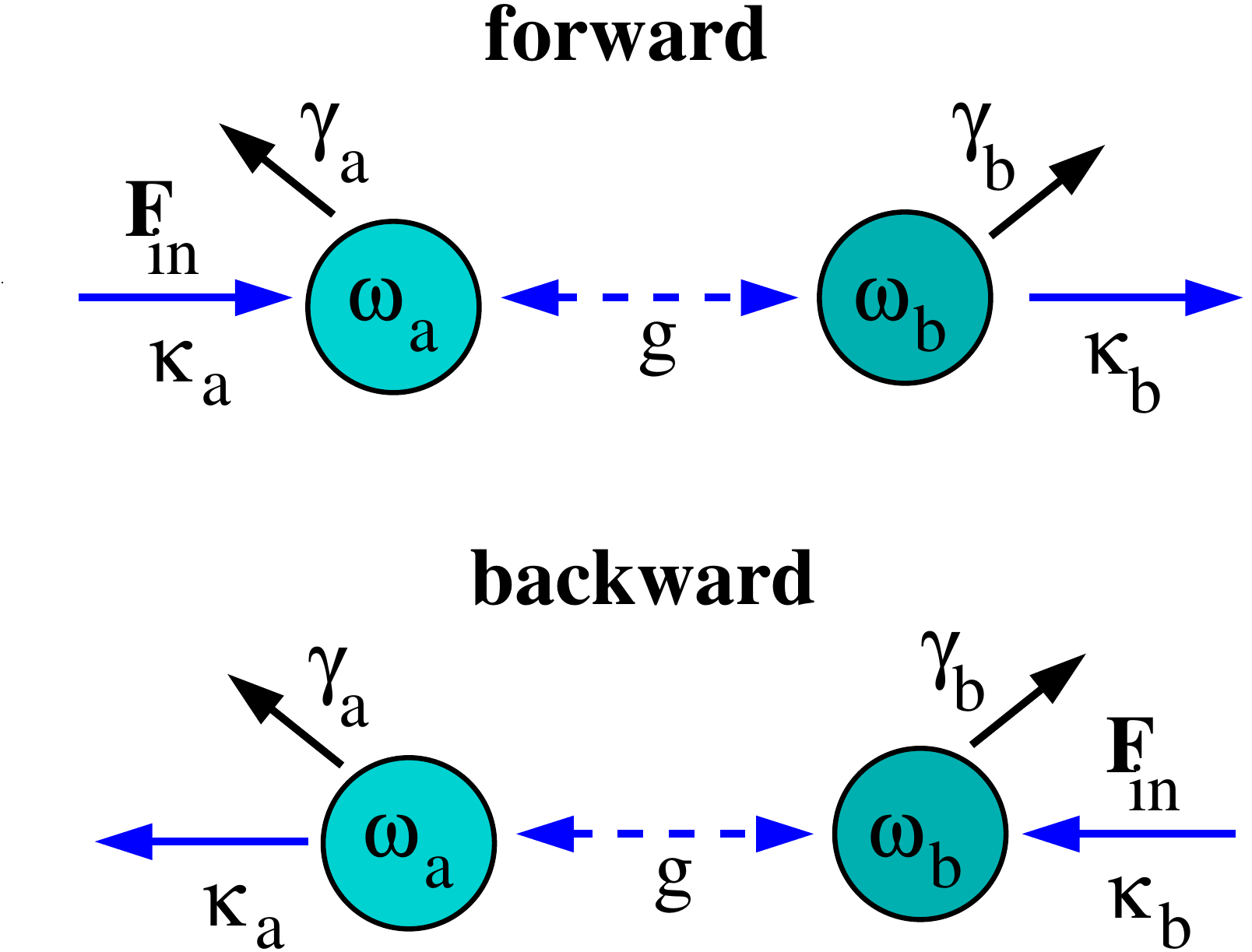}
	\caption{\label{Fig:NCcouplingFwdBwd} Sketch of the forward and backward transmission of an incident field $\mathbf{F}_{\rm in}$ in a dispersive coupling scenario.}
\end{figure}

The dynamical equations for two dispersively coupled resonators as depicted in Fig.~\ref{Fig:NCcouplingFwdBwd} are in principle well known but can also be derived from the above master equation in the case where the dissipative couplings $\kappa_{12}$ and $\kappa_{21}$ vanish which in our waveguide configuration happens when the distance between the resonators is chosen such that $\phi_{12} = \pi(n + 1/2)$ with integer $n$. Then taking $\Gamma_{12} = \Gamma_{21} \equiv g$ the dynamics is governed by (\ref{Eq:coupledOscillators}) but with the dynamical matrix in the rotating frame
\begin{equation}
	\tilde{\mathds{A}}^{\rm nc} =   \begin{pmatrix}
		-\ri \delta_{a,l} - \gamma_a -\kappa_a & \ri g \\
				    \ri g & -\ri \delta_{b,l} - \gamma_b - \kappa_b 
				  \end{pmatrix}
\end{equation}
where the dissipative terms $\gamma_a$ and $\gamma_b$ include only the internal and radiative losses, since there is no waveguide in this case, and $\kappa_a$ and $\kappa_b$ describe the coupling to input or output ports. In the following we will refer to this system as one with dispersive coupling using the abbreviation (nc) for non-dissipative coupling to distinguish it from the dissipative coupling. Similarly, we obtain the same Eq.~(\ref{Eq:CoupledFourier}) in Fourier space but with $\mathds{M}^{\rm nc}$ replaced by
\begin{equation}
	\mathds{M}^{\rm nc} =  \frac{1}{X_a^{\rm nc} X_b^{\rm nc} + g^2}\begin{pmatrix} X_b^{\rm nc} & \ri g \\ \ri g & X_a^{\rm nc} \end{pmatrix}
\end{equation}
and
\begin{equation}
	X_{a/b}^{\rm nc} = \ri \delta_{a/b} + \gamma_{a/b} + \kappa_{a/b}.
\end{equation}
The eigenvalues of the matrix $\mathds{M}^{\rm nc}$ are $\lambda_{\pm}^{\rm nc} = (X^{\rm nc} \pm g)^{-1}$.
Then in the limit $\gamma_{a/b} = 0$ and $\kappa_{a/b} = \Gamma$ with zero detuning $\delta_a, \delta_b = 0$ they are $\lambda_\pm^{\rm dc} \rightarrow (\ri \Gamma \pm g)^{-1}$. Hence, there is no BIC.

%In contrast, for the dispersive coupling in the corresponding limit $\gamma_{a/b} = \Gamma$ with zero detuning $\delta_a, \delta_b = 0$ we find $\lambda_\pm^{\rm dc} \rightarrow (\ri \Gamma \pm g)^{-1}$. In this case, there is not any particular strong response for any finite coupling and the linewidth of the response will be proportional to $\Gamma$.

\section{Synthetic fields and dissipatively coupled systems}

To introduce electric and magnetic synthetic fields we apply a modulation of both resonance frequencies of the form
\begin{align}
	\omega_{a} &\rightarrow \omega_a + \beta \cos(\Omega t),  \label{Eq:omegaa}\\
	\omega_{b} &\rightarrow \omega_b + \beta \cos(\Omega t + \theta) \label{Eq:omegab} 
\end{align}
with modulation frequency $\Omega$, modulation strength $\beta$, and a phase shift $\theta$. Then the coupled equations in Eq.~({\ref{Eq:CoupledFourier}) in Fourier space can be written as
\begin{equation}
	\boldsymbol{\psi} = \mathds{M} \tilde{\mathbf{F}} + \frac{\beta}{2 \ri} \mathds{M} \mathds{Q}_+ \boldsymbol{\psi}_+ + \frac{\beta}{2 \ri} \mathds{M} \mathds{Q}_- \boldsymbol{\psi}_- 
\label{Eq:coupledLangevin}
\end{equation}
with $\boldsymbol{\psi} = \bigl( a(\omega), b(\omega) \bigr)^t$, $\boldsymbol{\psi}_\pm = \bigl( a(\omega \pm \Omega),  b(\omega \pm \Omega) \bigr)^t$, and
\begin{equation}
	\mathds{Q}_\pm = \begin{pmatrix} 1 & 0 \\ 0 & \re^{\pm \ri \theta} \end{pmatrix}.
\end{equation}
Again, depending on dispersive or dissipative coupling $\mathds{M} = \mathds{M}^{\rm dc}$ or $\mathds{M} = \mathds{M}^{\rm nc}$, resp. The structure of the infinitely large set of equations nicely shows that due to modulation sidebands at $\omega_{a/b} \pm n \Omega$ with integer $n$ appear. This index can be understood as the coordinate of a synthetic dimension with an applied uniform synthetic electric field generating the sideband splitting~\cite{Lustig}. The above set of equations is a Floquet composition as encountered in standard Floquet approaches~\cite{Shirley}. It can be observed that due to the phaseshift $\theta$ the up- and downward transitions $\pm \Omega$ in this synthetic dimension introduce a phase factor $\exp(\pm \ri \theta)$ which can be understood as being induced by a synthetic magnetic field~\cite{Lustig}. This synthetic magnetic field is the reason for nonreciprocal energy transport~\cite{PetersonEtAl2019} and the breakdown of detailed balance for heat transport~\cite{SABGSA2023}.

To quantify such nonreciprocities we define transmission coefficients for a forward and backward transmission. 
For the dissipative coupling as depicted in Fig.~\ref{Fig:DCcoupling} in the forward case again the oscillator $a$ is driven by a monochromatic field $\mathbf{F} = (F_a,0)^t$ with $F_a = \epsilon \exp(-\ri \omega_l t)$ but due to the dissipative coupling by a waveguide, for instance, the transmission into the dissipative part of the system is considered so that
\begin{equation}
	\tau_{f}^{\rm dc} = \frac{\Gamma [a(t) + b(t)]}{F_a}.
\end{equation}
Likewise, in the backward case the oscillator $b$ is driven by a monochromatic field $\mathbf{F}_b = (0,F_b)^t$ with $F_b = \epsilon \exp(-\ri \omega_l t)$ so that the transmission is
\begin{equation}
	\tau_{b}^{\rm dc} = \frac{\Gamma [a(t) + b(t)]}{F_b}.
\end{equation}
On the other hand for the dispersive coupling as depicted in Fig.~\ref{Fig:NCcouplingFwdBwd} we define the forward transmission by assuming that the oscillator $a$ is driven by a monochromatic field $\mathbf{F}  = (F_a,0)^t$ with $F_a = \epsilon \exp(-\ri \omega_l t)$ and by quantifying the transmission through the coupled system by
\begin{equation}
	\tau_{f}^{\rm nc} = \frac{2 \kappa_b b(t)}{F_a}.
\end{equation}
In the backward case the oscillator $b$ is driven by a monochromatic field $\mathbf{F} = (0,F_b)^t$ with $F_b = \epsilon \exp(-\ri \omega_l t)$ so that the transmission through the coupled system is
\begin{equation}
	\tau_{b}^{\rm nc} = \frac{2 \kappa_a a(t)}{F_b}.
\end{equation}
These transmission coefficents are equivalent to those in Ref.~\cite{PetersonEtAl2019}. In order to evaluate these transmission coefficients we need to solve the set of coupled equations (\ref{Eq:coupledLangevin}).

To solve the set of equations in frequency space, first we choose a perturbation ansatz. The zeroth-order term 
is then
\begin{equation}
	\boldsymbol{\psi}^{(0)} = \mathds{M} \tilde{\mathbf{F}}
\end{equation}
which gives
\begin{align}
	\boldsymbol{\psi}^{(0)}_\pm &= \mathds{M}_\pm \tilde{\mathbf{F}}_\pm, \\
	\boldsymbol{\psi}^{(0)}_{++/--} &= \mathds{M}_{++/--} \tilde{\mathbf{F}}_{++/--}
\end{align}
for the frequencies $\omega \pm \Omega$ and $\omega \pm 2 \Omega$. The first perturbation correction is then
\begin{equation}
\begin{split}
	\boldsymbol{\psi}^{(1)} &= \frac{\beta}{2 \ri} \mathds{M} \mathds{Q}_+ \boldsymbol{\psi}_+^{(0)} + \frac{\beta}{2 \ri} \mathds{M} \mathds{Q}_- \boldsymbol{\psi}_-^{(0)} \\
				&= \frac{\beta}{2 \ri} \mathds{M} \mathds{Q}_+ \mathds{M}_+ \tilde{\mathbf{F}}_+ + \frac{\beta}{2 \ri} \mathds{M} \mathds{Q}_- \mathds{M}_- \tilde{\mathbf{F}}_-
\end{split}
\end{equation}
and the second order 
\begin{equation}
	\boldsymbol{\psi}^{(2)} = \frac{\beta}{2 \ri} \mathds{M} \mathds{Q}_+ \boldsymbol{\psi}_+^{(1)} + \frac{\beta}{2 \ri} \mathds{M} \mathds{Q}_- \boldsymbol{\psi}_-^{(1)}
\end{equation}
etc. To get the second-order term we have to determine the first order terms $\boldsymbol{\psi}^{(1)}$
at frequencies  $\omega \pm \Omega$ given by
\begin{align}
	\boldsymbol{\psi}^{(1)}_+ &= \frac{\beta}{2 \ri} \mathds{M}_+ \mathds{Q}_+ \boldsymbol{\psi}_{++}^{(0)} + \frac{\beta}{2 \ri} \mathds{M}_+ \mathds{Q}_- \boldsymbol{\psi}^{(0)} \\
				  &=  \frac{\beta}{2 \ri} \mathds{M}_+ \mathds{Q}_+ \mathds{M}_{++} \tilde{\mathbf{F}}_{++} + \frac{\beta}{2 \ri} \mathds{M}_+ \mathds{Q}_- \mathds{M} \tilde{\mathbf{F}} \\
	\boldsymbol{\psi}^{(1)}_- &= \frac{\beta}{2 \ri} \mathds{M}_- \mathds{Q}_+ \boldsymbol{\psi}^{(0)} + \frac{\beta}{2 \ri} \mathds{M}_- \mathds{Q}_- \boldsymbol{\psi}^{(0)}_{--} \\
				  &=\frac{\beta}{2 \ri} \mathds{M}_- \mathds{Q}_+ \mathds{M} \tilde{\mathbf{F}} + \frac{\beta}{2 \ri} \mathds{M}_- \mathds{Q}_- \mathds{M}_{--} \tilde{\mathbf{F}}_{--}.
\end{align}
Therefore, up to second order we obtain 
\begin{equation}
\begin{split}
	\boldsymbol{\psi}^{(2)} &= \mathds{M} \tilde{\mathbf{F}} \\
	  & \quad + \frac{\beta}{2 \ri} \mathds{M} \biggl[ \mathds{Q}_+ \mathds{M}_+ \tilde{\mathbf{F}}_+ + \mathds{Q}_- \mathds{M}_- \tilde{\mathbf{F}}_- \biggr] \\
			  &\quad -\frac{\beta^2}{4} \mathds{M} \biggl[\mathds{Q}_+  \mathds{M}_+ \mathds{Q}_- + \mathds{Q}_- \mathds{M}_- \mathds{Q}_+ \biggr] \mathds{M} \tilde{\mathbf{F}} \\
					  &\quad -\frac{\beta^2}{4} \mathds{M} \biggl[   \mathds{Q}_+  \mathds{M}_+ \mathds{Q}_+ \mathds{M}_{++} \tilde{\mathbf{F}}_{++}  \\
			  &\qquad \qquad+   \mathds{Q}_-  \mathds{M}_- \mathds{Q}_- \mathds{M}_{--} \tilde{\mathbf{F}}_{--} \biggr].
\end{split}
\end{equation}
For the evaluation of the transmission coefficients, we are only interested in the wave scattering of an incidenct monochromatic driving field so that only the terms containing the driving field $\tilde{\mathbf{F}}$ are relevant. Therefore we restrict ourselves to 
\begin{equation}
	\boldsymbol{\psi} = \mathds{M} \mathbf{F} - \frac{\beta^2}{4} \bigl[ \mathds{S} + \mathds{T}\bigr] \tilde{\mathbf{F}}
\end{equation}
with $\mathds{S} = \mathds{M} \mathds{Q}_+ \mathds{M}_+ \mathds{Q}_- \mathds{M}$ and $	\mathds{T} = \mathds{M} \mathds{Q}_- \mathds{M}_- \mathds{Q}_+ \mathds{M}$. As already noted in Ref.~\cite{PetersonEtAl2019} a second-order perturbation theory is needed to see any nonreciprocity, because the first order terms vanish. It can also be seen that the second order result is a superposition of a transmission paths along the upper and the lower sidebands.  

With these perturbative expressions we can derive the analytical formulas for the transmission coefficients in second-order. To simplify these expression we consider in the following only the case of oscillators with identical parameters $\omega_a = \omega_b = \omega_0$ and $\gamma_a + \kappa_a = \gamma_b + \kappa_b \equiv \gamma$. In the case of dissipative coupling we obtain 
\begin{equation}
\begin{split}
	\tau_{f/b}^{\rm dc} &=  \tau_0^{\rm dc} \biggl[ 1 -  \frac{\beta^2}{4} \bigl[ \lambda_+ (M_{11}^{{\rm dc},+} + M_{11}^{{\rm dc},-})\\
		   &\qquad + \cos(\theta) \lambda_+ (M_{11}^{{\rm dc},+} + M_{12}^{{\rm dc},-}) \\
		   &\qquad \mp \ri \sin(\theta) \lambda_- (M_{12}^{{\rm dc},-} - M_{12}^{{\rm dc},+}) \bigr] \biggr]
\end{split}
\label{Eq:pertsecondorderDC}
\end{equation}
with $\tau_0^{\rm dc} = \Gamma (M_{21}^{\rm dc} + M_{11}^{\rm dc}) = \Gamma \lambda_+^{\rm dc}$, $\lambda_-^{\rm dc} = {M_{11}^{\rm dc} - M_{12}^{\rm dc}}$ where all quantities need to be evaluated at $\omega = \omega_l$. Note, that $\lambda_\pm^{\rm dc}$ are the eigenvalues of the matrix $\mathds{M}^{\rm dc}$ explicitely given after Eq.~(\ref{Eq9}). As expected there can only be nonreciprocal transmission when a synthetic magnetic field is present, i.e.\ for $\theta \neq 0$. The largest nonreciprocal effect can be expected for $\theta = \pi/2 \pm n \pi$ with integer $n$. Focusing again to the case of identical resonators, due to the fact that $\lambda_-^{\rm dc}$ is the eigenvalue for the BIC having large values for small damping rates $\gamma$ we can expect very strong nonreciprocity for small $\gamma$ at a detuning $\delta = \omega_0 - \omega_l = 0$. Likewise, the factor $ M_{12}^{{\rm dc},-} - M_{12}^{{\rm dc},+}$ is for $\gamma \rightarrow 0$ given by
\begin{equation}
\begin{split}
	M_{12}^{{\rm dc},-} - M_{12}^{{\rm dc},+} &=  \Gamma\biggl( \frac{1}{(\delta + \Omega)(2 \ri \Gamma - (\delta + \Omega))} \\
	&\qquad- \frac{1}{(\delta - \Omega)(2 \ri \Gamma - (\delta - \Omega))} \biggr)
\end{split}
\end{equation}
indicating also strong nonreciprocity at the first two side-bands $\delta = \pm \Omega$. Taking higher orders into account we can expect to find strong reciprocity at $\delta = 0, \pm n \Omega$ with integer $n$. The modulation strength defines how strong the higher side bands contribute.

For the dispersive coupling we obtain
\begin{equation}
\begin{split}
	\tau_{f/b}^{\rm nc} &=   \biggl[\tau_0^{\rm nc} -  \frac{\beta^2}{2} \bigl[ \tau_0^{\rm nc} M_{11}^{\rm nc} (M_{11}^{{\rm nc},+} + M_{11}^{{\rm nc},-})\\
			    &\qquad + \cos(\theta) \lambda_1 (M_{12}^{{\rm nc},+} + M_{12}^{{\rm nc},-}) \\
			    &\qquad \mp \ri \sin(\theta) \lambda_2 (M_{12}^{{\rm nc},+} - M_{12}^{{\rm nc},-}) \bigr] \biggr]
\end{split}
\label{Eq:pertsecondorderNC}
\end{equation}
with $\tau_0^{\rm nc} = M_{21}^{\rm nc}$, $\lambda_1  = (M_{11}^{\rm nc})^2 + (M_{12}^{\rm nc})^2$, $\lambda_2 = (M_{12}^{\rm nc})^2 - (M_{11}^{\rm nc})^2$ where again all quantities are evaluated at $\omega = \omega_l$. As for the dissipative coupling we find nonreciprocity when a synthetic magnetic field is present.

\section{Enhanced nonreciprocity via synthetic magnetic fields --- Numerical results} 

A numerical solution of the coupled Langevin Eqs.~(\ref{Eq:coupledLangevin}) can be obtained by employing the same Floquet theory method as in Refs.~\cite{Shirley,PetersonEtAl2019,SABGSA2023}. By introducing the block vectors
\begin{align}
	\uline{\boldsymbol{\psi}} &= (\ldots, \boldsymbol{\psi}_{++},\boldsymbol{\psi}_+, \boldsymbol{\psi}, \boldsymbol{\psi}_-, \boldsymbol{\psi}_{--} \ldots)^t \\
	\uline{\mathbf{F}}  &=  (\ldots, \mathbf{F}_{++}, \mathbf{F}_+, \mathbf{F}, \mathbf{F}_-, \mathbf{F}_{--}, \ldots)^t
\end{align}
and the diagonal block matrix
\begin{equation}
			\uline{\mathds{M}} = \begin{pmatrix} 
					\ldots & \ldots & \ldots & \ldots & \ldots \\
					\ldots & \mathds{M}_{+} & \mathds{O} & \mathds{O} &\ldots \\
					\ldots &\mathds{O} & \mathds{M} & \mathds{O} &\ldots \\
					\ldots &\mathds{O} & \mathds{O} & \mathds{M}_- &\ldots  \\
                               		\ldots &\ldots & \ldots & \ldots & \ldots
	                     \end{pmatrix}
\end{equation}
and tridiagonal block matrix
\begin{equation}
			\uline{\mathds{L}} = \begin{pmatrix} 
					\ldots & \ldots & \ldots & \ldots & \ldots \\
					\frac{\ri \beta}{2} \mathds{M}_+ \mathds{Q}_+ & \mathds{1} & \frac{\ri \beta}{2} \mathds{M}_+ \mathds{Q}_- & \mathds{O} & \ldots \\
					\ldots &\frac{\ri \beta}{2} \mathds{M} \mathds{Q}_+ & \mathds{1} & \frac{\ri \beta}{2} \mathds{M} \mathds{Q}_- & \ldots \\
					\ldots &\mathds{O} & \frac{\ri \beta}{2} \mathds{M}_- \mathds{Q}_+ & \mathds{1} &\frac{\ri \beta}{2} \mathds{M}_- \mathds{Q}_- \\
                               		\ldots &\ldots & \ldots & \ldots & \ldots
	                     \end{pmatrix}
\end{equation}
we can rewrite the coupled Langevin Eqs.~(\ref{Eq:coupledLangevin}) as a block matrix equation
\begin{equation}
	\uline{\mathds{L}} \uline{\boldsymbol{\psi}} = \uline{\mathds{M}} \uline{\mathbf{F}}.
\end{equation}
Hence
\begin{equation}
	\uline{\boldsymbol{\psi}} = \uline{\mathds{L}}^{-1} \uline{\mathds{M}} \uline{\mathbf{F}}.
	\label{Eq:PerturbationSeries}
\end{equation}
By considering only block vectors of $2n + 1$ subvectors with the corresponding block matrices of $(2n + 1)\times(2n + 1)$ submatrices we obtain the  results up to order $n$. Numerically, we can choose $n$ for any modulation strength $\beta$ and modulation frequency $\Omega$ so large that the numerical result practically coincides with the exact solution. 

\begin{figure} 
  \includegraphics[angle=0,scale=0.7]{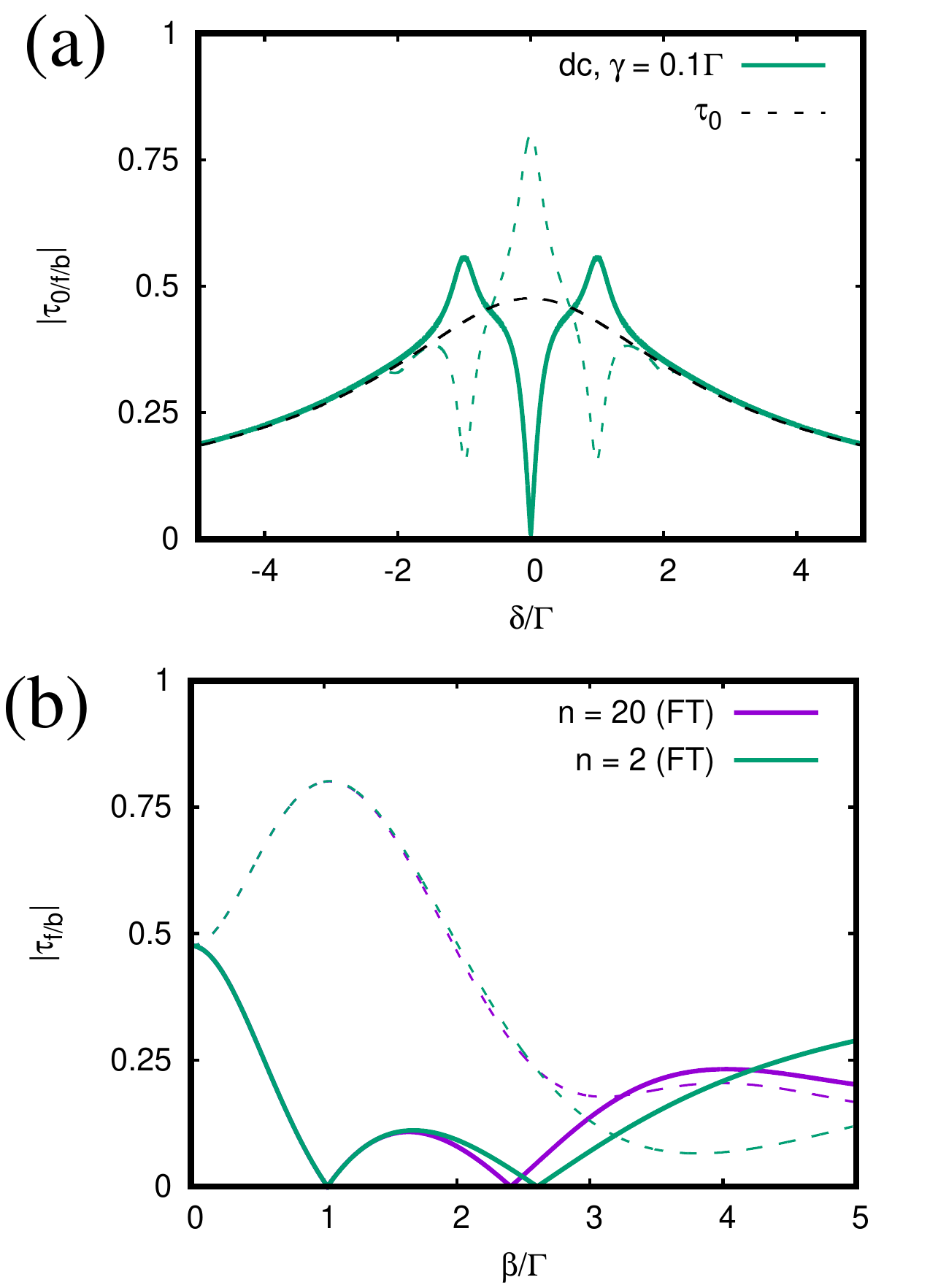}
	\caption{\label{Fig:ComparisionDCNC} Transmission for dissipative coupling (dc): (a) $|\tau_0^{\rm dc}|$ (dashed black line), $|\tau_f^{\rm dc}|$ (solid line) and $|\tau_b^{\rm dc}|$ (dashed line) using Floquet theory (FT) of n-th order ($n = 20$) with $\Omega = \Gamma$, $\beta = \Gamma$, $\gamma = 0.1\Gamma$ and $\theta = \pi/2$ as function of $\delta/\Gamma$ for the dissipative coupling. (b) Transmission as function of the modulation strength $\beta/\Gamma$ using the Floquet approach with $n = 2$ and $n = 20$ for $\delta = 0$ and $\theta = \pi/2$.} 
\end{figure}

\begin{figure}
  \includegraphics[angle=0,scale=0.7]{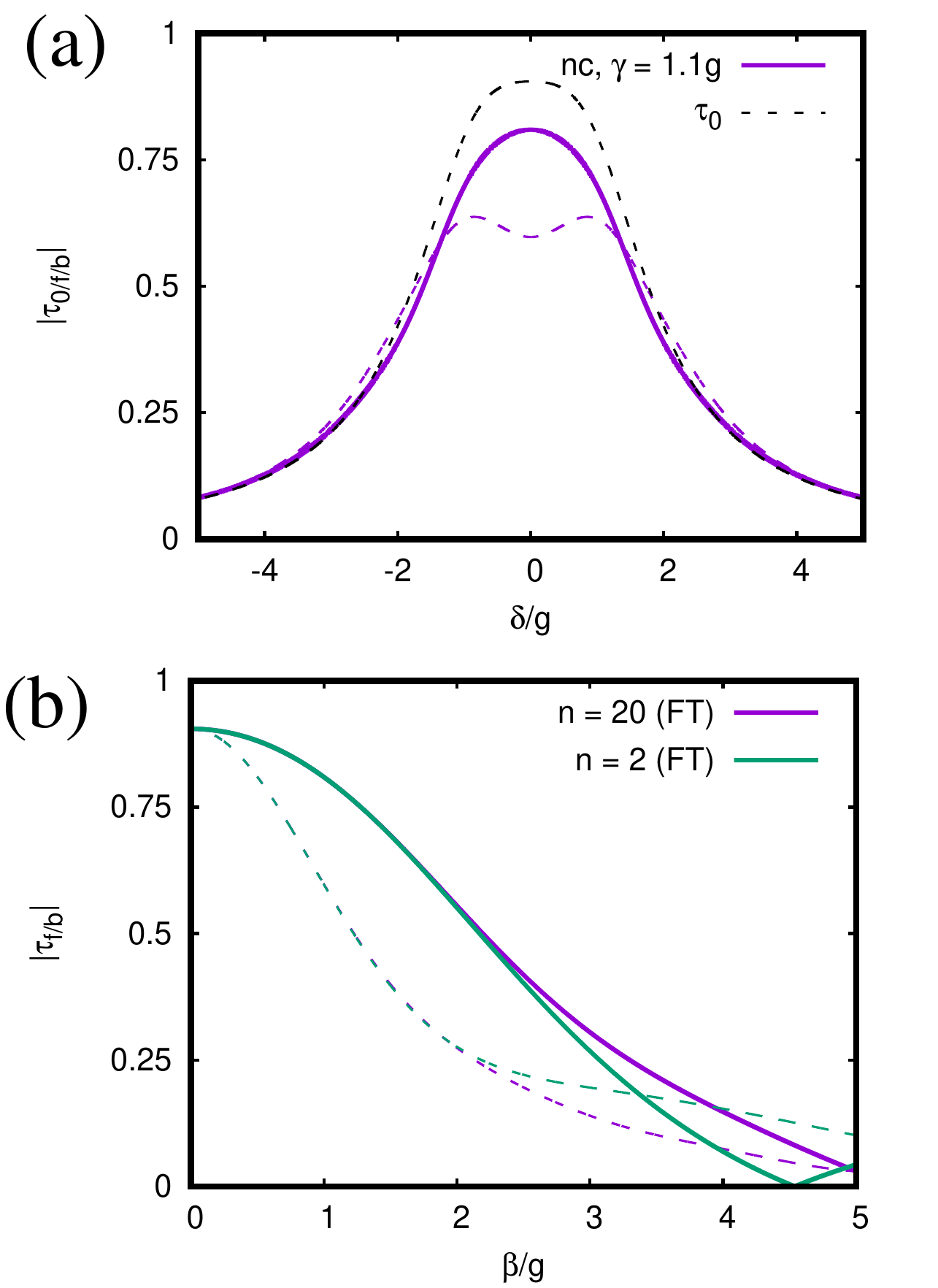}
	\caption{\label{Fig:ComparisionNC} Transmission for dispersive coupling (nc): (a) $|\tau_0^{\rm nc}|$ (dashed black line), $|\tau_f^{\rm nc}|$ (solid line) and $|\tau_b^{\rm nc}|$ (dashed line) using Floquet theory (FT) of n-th order ($n = 20$) with $\Omega = g$, $\beta = g$, $\gamma = 1.1g$ ($\kappa_a = \kappa_b = 1.0g$ and $\gamma_a = \gamma_b = 0.1g$), and $\theta = \pi/2$ as function of $\delta/g$ for the dissipative coupling. (b) Transmission as function  of the modulation strength $\beta/g$ using the Floquet approach with $n = 2$ and $n = 20$ for $\delta = 0$.} 
\end{figure}

In Fig.~\ref{Fig:ComparisionDCNC}(a) we show the modulus of the complex valued transmission coefficients $|\tau_{f/b}^{\rm dc}|$ for the forward and backward case for a dissipative coupling choosing the modulation parameters $\beta = \Gamma$, $\Omega = \Gamma$, $\theta = \pi/2$ and the damping rate $\gamma = 0.1\Gamma$. Note, that the transmissivity is defined by $|\tau_{f/b}^{\rm dc}|^2$. It can be observed that compared to the transmission without modulation $\tau_0^{\rm dc}$ the transmission in forward and backward direction can be enhanced or inhibited in forward or backward direction by turning on the modulation. Opposite trends at $\delta = 0,\pm \Omega$ can be observed, i.e.\ enhanced transmission at a frequency in forward direction is connected to an inhibited transmission in backward direction and vice versa. This trend can be understood from the last term in the analytical second order expressions in Eq.~(\ref{Eq:pertsecondorderDC}) which predict a symmetry between the enhancement and inhibition strength. However, the analytical second-order expressions in Eq.~(\ref{Eq:pertsecondorderDC}) only coincide with the numerical exact Floquet theory for $\beta/\Gamma \ll 1$ so that any asymmetry in the inhibition and enhancement around $\tau_0^{\rm dc}$ can be associated to higher order contributions in the perturbation approach. Interestingly, the Floquet approach in Eq.~(\ref{Eq:PerturbationSeries}) for $n = 2$ gives already a very good agreement with the numerical exact results for all $\beta/\Gamma \in [0:2]$. Furthermore, in Fig.~\ref{Fig:ComparisionDCNC} it can be seen that the forward transmission can be fully inhibited for $\beta/\Gamma \approx 1$ and $\beta/\Gamma \approx 2.4$ at zero detuning $\delta = 0$. For $\beta/\Gamma \approx 1$ the difference in forward and backward transmission reaches a maximum when choosing $\theta = \pi/2$.

\begin{figure}%[!hbt]
  \includegraphics[angle=0,scale=0.7]{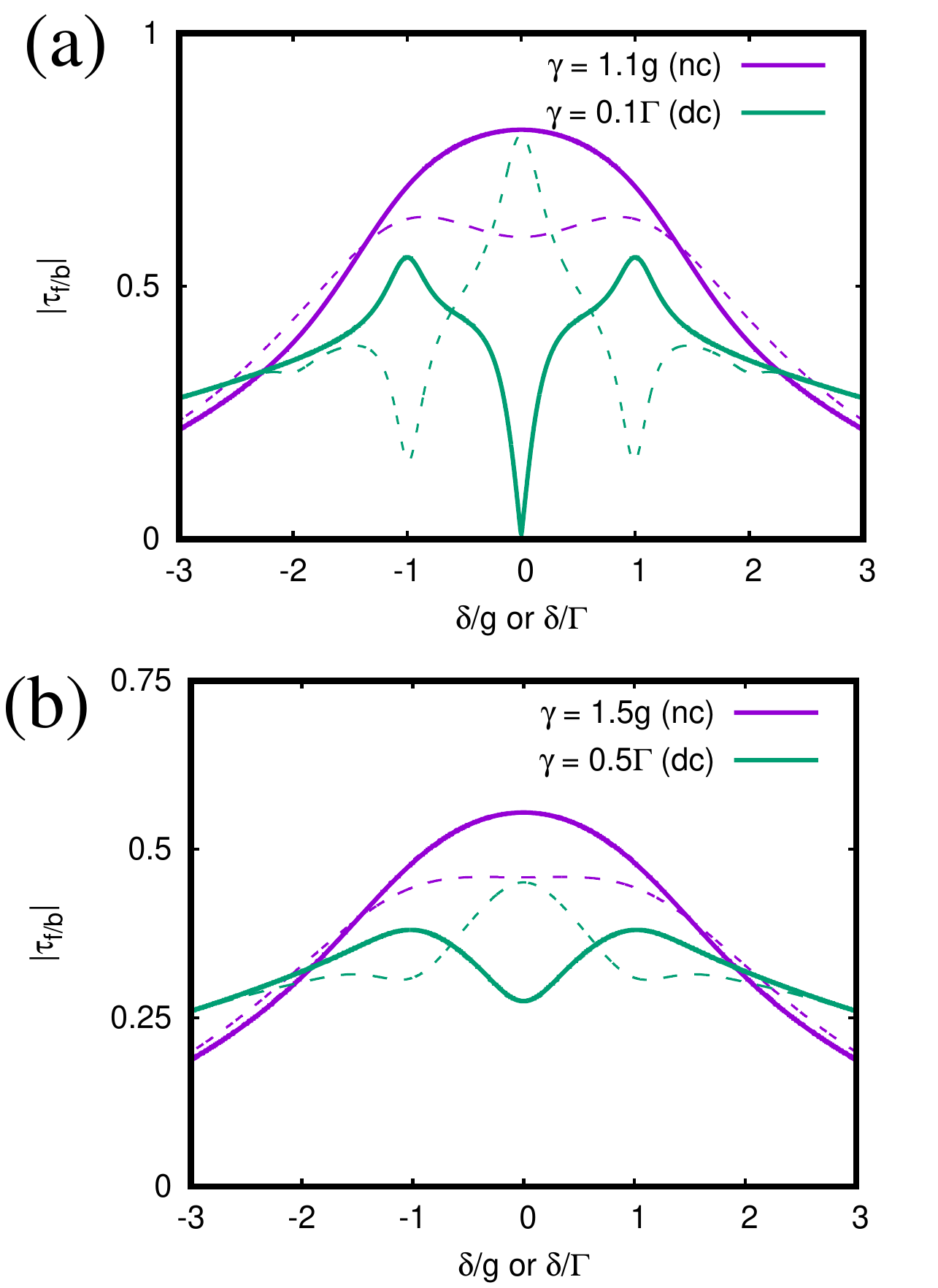}
	\caption{\label{Fig:DCNCcomparison} Comparison of dissipative coupling (dc) and dispersive coupling (nc): $|\tau_f^{\rm nc/dc}|$ (solid line) and $|\tau_b^{\rm nc/dc}|$ (dashed line) using Floquet theory of n-th order ($n = 20$) with $\Omega = \Gamma$, $\beta = \Gamma$ and (a) $\gamma = 0.1\Gamma$ and (b) $\gamma = 0.5\Gamma$ as function of $\delta/\Gamma$ for the dissipative coupling and $\Omega = g$, $\beta = g$ with overall damping $\gamma = 1.1 g$ and  $\gamma = 1.5 g$ as function of $\delta/g$ for the dispersive coupling. In both cases $\theta = \pi/2$.}
\end{figure}

We compare now the dissipative coupling with the dispersive coupling by determining the transmission coefficients $|\tau_{f/b}^{\rm nc}|$ for the choice of parameters corresponding to the dissipative coupling, i.e.\ we chose for the modulation strength and frequency $\beta = g$ and $\Omega = g$ and for the damping rate $\gamma = 1.1 g$ ($\kappa_a = \kappa_b = 1.0g$ and $\gamma_a = \gamma_b = 0.1g$) which corresponds to the overall damping $1.1 \Gamma$ in the dissipative coupling scenario. In Fig.~\ref{Fig:ComparisionNC}(a) it can be seen that for the chosen parameters due to modulation there is always an inhibition of the transmission compared with the unmodulated transmission $|\tau_{0}^{\rm nc}|$. This inhibition at zero detuning $\delta = 0$ is for small enough modulation strength $\beta/g$ much stronger for the backward than for the forward scenario as can also be seen in Fig.~\ref{Fig:ComparisionNC}(b). With our choice of parameters the nonreciprocal transmission at the side-bands at $\delta = \pm \Omega$ cannot be seen. Both observations are in stark contrast to the dissipative coupling where the transmission can exceed $|\tau_{0}^{\rm dc}|$ and the nonreciprocal transmission at the side-bands $\delta = \pm \Omega$ is clearly visible.

\begin{figure}
 \includegraphics[angle=0,scale=0.7]{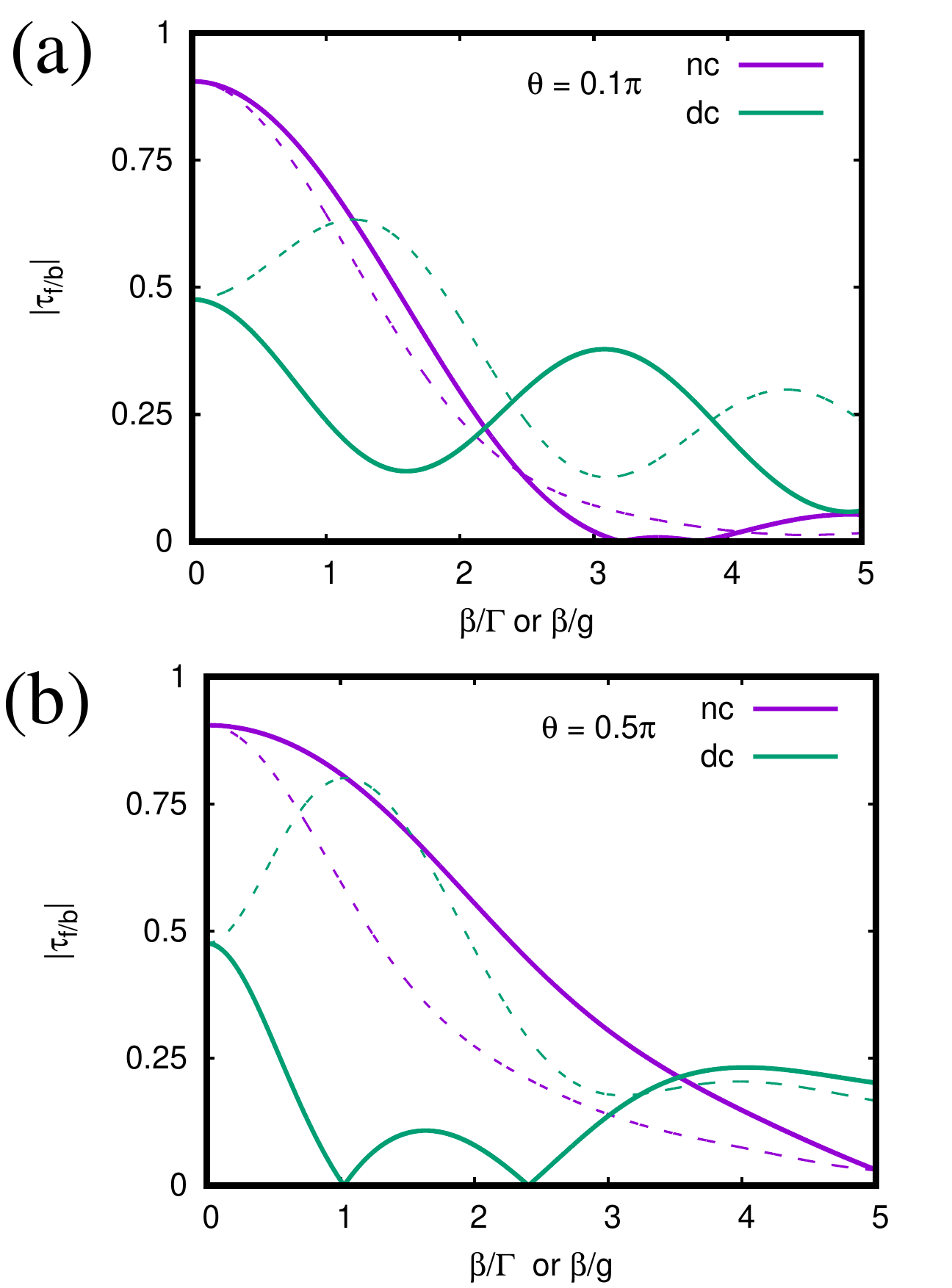}
	\caption{\label{Fig:Diffangles} Comparison of dissipative coupling (dc) and dispersive coupling (nc):  $|\tau_f|$ (solid line) and $|\tau_b|$ (dashed line) using perturbation theory of n-th order ($n = 20$) with $\Omega = \Gamma$, $\beta = \Gamma$ and $\gamma = 0.1\Gamma$ so that the overall damping is $1.1\Gamma$ for the dissipative coupling and $\Omega = g$, $\beta = g$ with overall damping $\gamma = 1.1 g$ as function of $\beta/g$ for the dispersive coupling for different relative dephasing $\theta = 0.1\pi$ and $\theta = 0.5\pi$ with $\delta = 0$.}
\end{figure}

To more clearly demonstrate the difference between the transmissions in the dispersive and dissipative case we show in Fig.~\ref{Fig:DCNCcomparison} a direct comparison for two different values for the damping rates. It can be seen that for small damping rates the nonreciprocal effect becomes stronger as well as the difference between the dispersive and dissipative case. Furthermore, at zero detuning the roles of forward and backward transmission are interchanged in both cases. In the dissipative case the backward transmission is enhanced with respect to $|\tau_0^{\rm dc}|$ and the forward transmission is inhibited whereas in the dispersive case it is the forward transmission which is stronger than the backward transmission. However, if the forward or backward transmission dominates depends on the modulation strength and the phase $\theta$. In Fig.~\ref{Fig:Diffangles} the dependence on the modulation strength for two different values of $\theta$ is shown. It can be observed that a complete inhibition for both the dispersive and dissipative coupling is for certain values of $\theta$ and $\beta$ possible. Furthermore the nonreciprocity for dissipative coupling is in nearly all cases stronger than for dispersive coupling. We further remark that for any phase $\theta$ we obtain the same results in both cases as for $-\theta$ but with the difference that the roles of forward and backward transmission are again interchanged. This property is also suggested from the second order expressions in Eqs.~(\ref{Eq:pertsecondorderDC}) and (\ref{Eq:pertsecondorderNC}). Hence, by choosing either positive or negative values of $\theta$ one can actively control the direction of preferred transmission.

\section{Line narrowing in synthetic magnetic fields}

Finally, we want discuss another important difference between the dissipative and dispersive coupling which are the different linewidths observed at $\delta = 0, \pm \Omega$ (see Fig.~\ref{Fig:DCNCcomparison}, for example). For the dissipative coupling the zeroth-order linewidth is determined by the transmission coefficient
\begin{equation}
	\tau_0 = \Gamma (M_{21}^{\rm dc} + M_{11}^{\rm dc}) = \frac{\Gamma}{\ri \delta + (\gamma + 2\Gamma)} 
\end{equation}
so that the linewidth of $|\tau_0|$ or $|\tau_0|^2$ is 
\begin{equation}
	\Delta_{\rm dc}^{(0)} = 2 (\gamma + 2\Gamma)
\end{equation}
and in the limit $\gamma \ll \Gamma$ we have
\begin{equation}
	\Delta_{\rm dc}^{(0)} = 4 \Gamma.
\end{equation}
In second order we can determine the linewidth by considering the nonreciprocal part of the transmission coefficients in Eq.~(\ref{Eq:pertsecondorderNC}) by determining $\tau_0 \lambda_-^{\rm dc}$ for the resonances at $\delta = 0$ and $M_{12}^{{\rm dc},+} - M_{12}^{{\rm dc},-}$ for the resonances at $\delta = \pm \Omega$. First, we compute 
\begin{equation}
\begin{split}
	\tau_0 \lambda_-^{\rm dc} &= \frac{\Gamma}{X^2 - \Gamma^2} \\
			 &= \frac{\Gamma}{(\ri \delta + \gamma + \Gamma)^2 - \Gamma^2} \\
			 &= \frac{\Gamma}{\delta(2 \ri (\gamma + \Gamma) - \delta) + \gamma^2 + 2\gamma \Gamma}.
\end{split}
\end{equation}
Close to $\delta \approx 0$ [$\delta \ll 2 (\gamma + \Gamma)$] we therefore have
\begin{equation}
	\tau_0 \lambda_-^{\rm dc} \approx  \frac{\Gamma}{2 \ri (\gamma + \Gamma)}\frac{1}{\delta  - \ri \frac{\gamma^2 + 2\gamma\Gamma}{2 (\gamma + \Gamma)}}
\end{equation}
so that we can approximate the linewidth by
\begin{equation}
	\Delta_{\rm dc}^{(2)} \approx \frac{\gamma^2 + 2\gamma \Gamma}{\gamma + \Gamma} =  \gamma\frac{\frac{\gamma}{\Gamma} + 2}{\frac{\gamma}{\Gamma} + 1}.
\end{equation}
We obtain the same approximative expressions for the resonances at $\delta = \pm \Omega$. In the limit $\gamma \ll \Gamma$ we obtain
\begin{equation}
	\Delta_{\rm dc}^{(2)} \approx 2 \gamma.
\end{equation}
Therefore, in the limit $\gamma \ll \Gamma$ where the BIC becomes relevant the linewidth of the zeroth order corresponds to that of $\lambda_+^{\rm dc}$ and the linewidth at the resonances $\delta = 0, \pm \Omega$ have the linewidth of the BIC $\lambda_-^{\rm dc}$. 

Let us now turn to the linewidth for the dispersive coupling. The zeroth-order transmission $|\tau_0^{\rm nc}|$ or $| \tau_0^{\rm nc}|^2$ can be derived from
\begin{equation}
	\tau_0^{\rm nc} = \frac{2 \ri g \kappa}{X^2_{\rm nc} + g^2} \approx  \frac{1}{\gamma}\frac{\ri g \kappa}{\delta \ri + \frac{\gamma^2 + g^2}{2 \gamma}}
\end{equation}
for $\delta \approx 0$. The approximative linewidth is therefore
\begin{equation}
	\Delta_{\rm nc}^{(0)} \approx \frac{\gamma^2 + g^2}{\gamma}.
\end{equation}
For the linewidth in second-order we can study the transmission coefficients in Eq.~(\ref{Eq:pertsecondorderNC}). The sine-term is the one which determines the nonreciprocal effect so that we can focus on this term. Therefore, we need to focus either on $\lambda_2$ for $\delta \approx 0$ or $M_{12}^+ - M_{12}^-$ for $\delta \approx \pm \Omega$. Now, since we have
\begin{equation}
	\lambda_1 = \frac{1}{X^2_{\rm nc} +g^2}
\end{equation}
and
\begin{equation}
	M_{12}^+ - M_{12}^- = \ri g \biggl( \frac{1}{X^2_{\rm nc,+} +g^2} - \frac{1}{X^2_{\rm nc,-} +g^2}  \biggr)
\end{equation}
it is apparent that both expressions have the same depencence at the resonances as the zeroth order term $\tau_0^{\rm nc}$ so that the linewidth of the nonreciprocal part at $\delta = 0, \pm \Omega$ is the same as for $\tau_0^{\rm nc}$.

Hence, by employing a synthetic field the linewidth is narrowing from approximately $4 \Gamma$ to $2 \gamma$ in the dissipative coupling scenario, whereas in the dispersive coupling the linewidth is not altered by the modulation. The linewidth narrowing can be nicely observed in Fig.~\ref{Fig:ComparisionDCNC}(a) and the difference between the dispersive and dissipative case is nicely observable in Fig.~\ref{Fig:DCNCcomparison}.

\section{Conclusions}

In summary, we have investigated the nonreciprocal transmission properties of two dissipatively coupled resonators due to 
the presence of synthetic electric and magnetic fields. Our results clearly show that in the dissipatively coupled system the forward and
backward transmission are either enhanced or reduced at detunings $\delta = 0, \pm \Omega, \ldots$ with respect to the 
transmission without frequency modulation. This is in contrast to the dispersive coupling where forward and backward
transmission are both reduced and the nonreciprocity consists in the fact that one direction is stronger reduced than 
the other. We further find that the difference between forward and backward transmission is much stronger in the dissipative
than in the dispersive case in particular when the intrinsic and radiative losses are minimized so that the BIC
dominates the nonreciprocal transmission effect. For certain values of the modulation strength $\beta$ we find for zero detuning $\delta = 0$ a zero transmission in one direction depending on the sign of the phase $\theta$ either in forward or backward direction.
Hence, the energy transfer is clearly uni-directional like for a diode. Such a diode effect can also be obtained for the 
dispersive coupling but for much larger values of the coupling strength and with a much smaller value of the 
transmission $|\tau^{\rm nc}| < 0.1$ in the non-blocked direction compared to $|\tau^{\rm dc}| > 0.75$. Finally, 
for $\gamma \ll \Gamma$ we find 
that the linewidths of nonreciprocal transmission at detunings $\delta = 0, \pm \Omega, \ldots$ are narrowing from $4 \Gamma$
without modulation to $2 \gamma$ with modulation in the case of dispersive coupling, whereas for the dissipatively coupled 
system the linewidths are the same with and without modulation.

\acknowledgements

S.-A.\ B.\ acknowledges support from Heisenberg Programme of the Deutsche Forschungsgemeinschaft (DFG, German Research Foundation) under the project No.\ 461632548 and he thanks the University of Montpellier and the group Theory of Light-Matter and Quantum Phenomena of the Laboratoire Charles Coulomb for hospitality during his stay in Montpellier where part of this work has been done. G.S.\ A.\ thanks the kind support of The Air Force Office of Scientific Research [AFOSR award no. FA9550-20-1-0366] and The Robert A. Welch Foundation [grant no. A-1943].

\end{document}